\documentclass[conference]{IEEEtran}
\IEEEoverridecommandlockouts

\usepackage{cite}
\usepackage{amsmath,amssymb,amsfonts}
\usepackage{algorithm}
\usepackage{algpseudocode}
\usepackage{graphicx}
\usepackage{textcomp}
\usepackage{hyperref}
\usepackage{xcolor}
\algrenewcommand\alglinenumber[1]{}
\def\BibTeX{{\rm B\kern-.05em{\sc i\kern-.025em b}\kern-.08em
    T\kern-.1667em\lower.7ex\hbox{E}\kern-.125emX}}
    
\begin{document}

\title{A customizable inexact subgraph matching algorithm for attributed graphs}

\author{\IEEEauthorblockN{Tatyana Benko}
\IEEEauthorblockA{ 
\textit{University of Oregon}\\
Eugene,  Oregon, USA
\\
tbenko@uoregon.edu}
\and
\IEEEauthorblockN{Rebecca Jones}
\IEEEauthorblockA{\textit{Pacific Northwest National Laboratory} \\
Richland, Washington, USA\\
rebecca.d.jones@pnnl.gov}
\and
\IEEEauthorblockN{ Lucas Tate}
\IEEEauthorblockA{\textit{Pacific Northwest National Laboratory} \\
Richland, Washington, USA \\
lucas.tate@pnnl.gov}

}

\maketitle

\begin{abstract}
Graphs provide a natural way to represent data by encoding information about objects and the relationships between them. With the ever-increasing amount of data collected and generated, locating specific patterns of relationships between objects in a graph is often required. Given a larger graph and a smaller graph, one may wish to identify instances of the smaller query graph in the larger target graph. This task is called subgraph identification or matching. Subgraph matching is helpful in areas such as bioinformatics, binary analysis, pattern recognition, and computer vision. In these applications, datasets frequently contain noise and errors, thus exact subgraph matching algorithms do not apply. In this paper we introduce a new customizable algorithm for inexact subgraph matching. Our algorithm utilizes node and edge attributes which are often present in real-world datasets to narrow down the search space. The algorithm is flexible in the type of subgraph matching it can perform and the types of datasets it can process by its use of a modifiable graph edit distance cost function for pairing nodes. We show its effectiveness on family trees graphs and control-flow graphs.

\end{abstract}

Many fields use graphs to model data consisting of objects and relationships between them, such as social networks \cite{ma2018comparative}, road networks \cite{unsalan2012road}, and biological networks \cite{saga}. Vertices of a graph represent objects in the network, and edges identify relationships between objects. Important information about each specific object and relationship may be incorporated into the graph as node and edge attributes. For example, in a graph of a social network, vertices represent people, and friends share an edge between them. The associated node attributes might include information such as age, hobbies, and interests of each person, and edge attributes can indicate the type of relationship, such as family, friends, or colleagues. Subgraph matching can be used in a social network graph to locate specific relationships between groups of people, such as families \cite{ma2018comparative}. \\

We define the subgraph matching task as follows. Given a query graph $G_Q = (V_Q, E_Q)$ and a target graph $G_T = (V_T, E_T)$, the subgraph matching task involves finding an injective map $f$ between the vertices of $G_T$ and the vertices of $G_Q$, such that the induced subgraph on the mapped vertices of $G_T$ is as close to $G_Q$ as possible. If we require that the subgraph of $G_T$ induced by the mapped nodes is isomorphic to $G_Q$, this is called exact subgraph matching. However, in many real-world datasets it's important to consider possible noise or mistakes in the graphs created from the data. Inexact subgraph matching addresses this issue by finding the best matching induced subgraph, which might not be isomorphic to the query graph. Different research areas define the “best” matching subgraph in different ways, thus inexact subgraph matching algorithms are more diverse than exact ones. The correctness of inexact mappings is often measured by a cost function which compares graph structure and node attributes between the graphs.\\

Inexact subgraph matching is a valuable tool across various fields. Say a researcher in bioinformatics is studying how a particular molecular regulator (such as DNA, RNA, protein, or any combination of these that form a complex) interacts with others; this can be formulated as an inexact subgraph matching problem \cite{gaddi}, \cite{carletti2015vf2}, \cite{bonnici2013subgraph}. The target graph is the entire gene regulatory network, and the query graph is the graph representation of the molecular regulator of interest. Locating each instance in the network allows the researcher to see where the interactions occur, and which molecular regulators are involved. In binary analysis, it is often useful but difficult to locate third-party libraries and function reuse in binary code \cite{libam} \cite{dullien2005graph}, \cite{tang2022libdb}, \cite{xu2021interpretation}. The difficulty of this task stems from the variability in binaries produced by different compilers and optimization levels. Inexact subgraph matching can be applied to this problem by searching for the graph representation of the library or function in the larger graph representing the binary. Subgraph matching can also aid tasks in pattern recognition and computer vison by identifying specific symbols or objects in a larger image or diagram \cite{conte2003graph}, \cite{llados2001symbol}, \cite{bunke1995efficient}, \cite{christmas1995structural}, \cite{conte2004thirty}. \\

In this paper, we propose a new algorithm for inexact subgraph matching. First, the algorithm finds potential starting nodes using node attributes and the graph structure of a small surrounding area. Next, it explores the target graph using depth-first search, pairing nodes of the target with nodes in the query graph as it explores. At each vertex, it evaluates each viable potential pairing using a graph edit distance cost function, which is adjustable depending on the user's needs (for example, matching node attributes correctly may be more important than matching all the edges exactly). If multiple matches exist, the algorithm finds these mappings by backtracking in the depth-first search. The user can specify whether inexact matchings should be found, or only exact matchings. \\

The main contribution of this paper is a highly customizable algorithm for inexact subgraph matching, which is also capable of exact matching. The modifiable cost function allows the user to set the level of importance of correct node attribute matching versus correct graph structure matching. This feature, along with our graph traversal approach, permits use of the algorithm on datasets with varying levels of node attribute quality, that is, how well the node attributes uniquely identify nodes. \\

In Section \ref{sec:existingMethods} we discuss existing algorithms for inexact subgraph matching. We describe our algorithm in Section \ref{sec:algorithm}, and discuss our experiments and results in Section \ref{sec:experiments}. Sections \ref{sec:conclusion} and \ref{sec:future_work} contain our conclusions and future work. \\

\section{Existing Methods}\label{sec:existingMethods}
The naive method for subgraph matching is a brute force tree search: take all subsets of vertices in your target graph of size $|V_Q|$, look at the induced subgraph in $G_T$, and compare it to $G_Q$. While this process would eventually produce the correct match, it very quickly becomes computationally infeasible as you increase the sizes of your graphs. In fact, the subgraph matching problem is known to be NP-complete \cite{cook2023complexity}. The brute-force tree search can be modified, however, by incorporating ways to greatly reduce the search space by ruling out incorrect mappings as early as possible. Ullmann proposed the first algorithm to do this for exact subgraph matching in 1976 \cite{ullmann1976algorithm}. He introduced a refinement process to the brute-force tree search, which checks the neighbors of potential node matches at each step in the process. Many exact graph matching algorithms used today are based on Ullmann's algorithm \cite{VF2}, \cite{VF3}, \cite{shang2008taming}, \cite{gaddi}, \cite{he2008graphs}, \cite{zhao2010graph}, each with additional novel approaches for reducing the search space. \\

The strategy for inexact subgraph matching differs from that of exact matching in several ways. We cannot rule out potential node matches based on a minor difference in degree, or node attributes, as we can in exact matching. Thus, there are more possible mappings to search through to find the best one. The next difference comes from the question, how exactly do we define the ``best" matching subgraph? Varying answers to this question coming from different fields results in more diversity in inexact subgraph matching algorithms. However, many existing algorithms still follow a similar overall strategy, while the differences between algorithms are found within the details of each step. First, some initial node assignments are made based on a comparison (for example, exact matching on node labels, or by indexing some feature of the graphs and pairing these first) and used as initial pairings. The graphs are then explored from these initial matches using a scoring function to measure ``correctness" of each further pairing made. Some backtracking and node re-assignments may be done in order to find the optimal mapping. Variations among existing methods occur within how the initial assignments are made, definition and type of pairing function, how data is stored and retrieved during the graph exploration, and how the backtracking is achieved.\\

Some existing algorithms that follow this broad approach of creating node pairings during graph exploration via a scoring system include G-Finder, \cite{gfinder}, SAGA \cite{saga}, TALE \cite{tale}, an algorithm by Tu et. al. \cite{tu2020inexact}, MAGE \cite{pienta2014mage}, and our algorithm. G-Finder creates a novel data structure which is used to find the optimal matching order on the nodes in the query graph. SAGA is designed for use on biological networks; it makes initial matches via an index created from substructures of vertices in the two graphs, and then combines these fragments to create a matching subgraph using a distance function. TALE is an algorithm designed to work efficiently on large target and query graphs. It uses a novel indexing method for initial node assignments such that the index size scales linearly with the graph database size. The query graph is explored from these initial nodes, using a cost function to find the optimal subgraph match. Tu et. al. introduce an algorithm which explores the query graph via depth-first search, and makes node assignments based on a graph edit distance function. They use constrained lower bounds on the graph edit distance function as a heuristic in the greedy depth-first search. MAGE makes initial node assignments based
on node attributes and centrality, and makes new assignments based on a random walk with restart proximity score between nodes. \\

While following this broad approach for inexact subgraph matching, our algorithm offers modifications within the graph traversal approach and the node pairing cost function, allowing for a high-degree of customizability throughout all steps of the algorithm. Some aspects that are easily set by the user include:

\begin{itemize}
    \item node attributes and string similarity comparison method used for finding initial node pairings,
    \item node and edge attributes used during the depth-first search for node matching,
    \item parameters in the cost function indicating level of importance of node attributes and graph structure,
    \item threshold value for acceptable matches during the depth-first search,
    \item exact versus inexact matching.
\end{itemize}

In addition to these parameters, our method also differs from existing ones in the graph traversal approach. For our graph traversal method, we use depth-first search. While this is not a novel approach, our algorithm is unique in that we traverse the target graph, instead of the query graph. The motivation for this is as follows. Although performing the depth-first search in the target graph means there are more nodes to visit, at each node there are significantly fewer nodes to compare. To make this a beneficial strategy, we limit the size of the target graph to be explored based on the start node and the size of the query graph (size meaning maximum path length between any two nodes). We note that in the case where the node attributes significantly narrow down the pairing options for each node, this may not be the best approach. However, if the node attributes do not uniquely identify node pairings (which is the case in all our experiments), or do not  narrow down the options enough, this offers a useful approach. \\

We note that worst case time complexity for our algorithm is $O(|V_T|^3 \cdot |V_Q|^2 + |V_T|^2 \cdot |V_Q|^2 \cdot |E_T|)$, comparable to that of G-Finder. This would happen if there were no node attributes to compare during the depth-first search, thus it compares every possible node. For G-Finder, the worst case time complexity happens when the target graph is a complete graph. In the best case for our algorithm, the time complexity is the same as that of G-Finder, $O(|V_Q|)$, occuring if the matching subgraph is found on the first start node and the first depth-first search with no backtracking.\\

\section{The Algorithm} \label{sec:algorithm}
\subsection{Overview}

Our algorithm performs a depth-first search in $G_T$, and matches explored nodes to nodes in $G_Q$ based on the cost of each node association (calculated via the cost function defined in Section \ref{sec:costFunctions}). Using node and edge attributes as well as graph structure, we refine the search space before and during the depth-first search. Backtracking in the depth-first search allows us to locate the best possible subgraph, or multiple matches if they exist. The algorithm is equipped for exact and inexact subgraph matching, and the cost functions for inexact matching are adjustable depending on the dataset and the type of matching desired. \\

\subsection{Finding Start Nodes}
One of the required inputs for the algorithm is a list of starting nodes for the depth-first search in the target graph, along with potential matches in the query graph. Ideally, the starting node matches are as exact as possible. In our experiments, we find potential start nodes using both node attributes and graph structure. First, matches are made for all the nodes in the target graph using a string similarity score to compare node attributes (we use Jaro-Winkler distance in our experiments). The node attributes used for this comparison are adjustable; the user may define whether to use specific node attributes or all of them, as well as set the threshold for the string similarity score. Next, we take the potential start nodes to be those with less than a user-set parameter $k$ of matches in the query graph, and do an exploration of a small area of the graph around each of these nodes to produce the final list of start nodes. In some cases, this may produce exact pairings of a node in the target graph with one node in the query graph, while in other cases there may be multiple potential matches in the query graph. If no start nodes are found in this process, one may increase the value of $k$ to obtain them. \\

The graph exploration component around each potential starting node $u \in V_T$ involves several degree checks (see Algorithm 1). If any of the degree checks comparing node $u$ to a potential match $w \in V_Q$ fail, it removes $w$ from the list of potential matches for $u$. For the first degree check, it evaluates whether $\operatorname{deg}(u) \geq \operatorname{deg}(w)$. If this fails, then it will not be possible to match all of the neighbors of node $w \in V_Q$ with neighbors of node $u \in V_T$, thus we would automatically have unmatched nodes in the query graph. Second, it compares the number of 2-hop neighbors of $u$ to that of $w$, where the set of 2-hop neighbors of  $u$ is the set of all neighbors of neighbors of $u$. Third, it checks for each neighbor and for each 2-hop neighbor of node $w$, if there is at least one potential match that is a neighbor or 2-hop neighbor of $u$ (where a potential match is given by a comparison of node attributes). \\


\begin{algorithm}[H]
    \caption{FindStartNodes}
    \begin{algorithmic}[1]
    \Require target graph $G_T$, 
    \State query graph $G_Q$,
    \State dictionary $P = \{u: \{w_1, w_2, ... \}\}$ for nodes $u \in V_T$ and nodes $w_i \in V_Q$ which are possible matches for $u$ based on node attributes, parameter $k$
    \State \textbf{Output:} dictionary $S$ of start nodes in $G_T$ and potential matches in $G_Q$
    \For {$u \in V_T$}
    \If {$u$ has less than $k$ possible matches as given by $P$}
    \State $S \gets u$
    \EndIf
    \EndFor
    \For {$u \in S$}
        \For {$w$ a possible match for $u$ as given by $P$}
            \If {$\operatorname{deg}(u) < \operatorname{deg}(w)$}
            \State continue
            \EndIf
            \If {number of 2-hop neighbors of $u < $ number of 2-hop neighbors of $w$}
            \State continue
            \EndIf
            \If {$\exists$ neighbor of $w$ that has no potential match as given by $P$}
            \State continue
            \EndIf
            \If {$\exists$ 2-hop neighbor of $w$ that has no potential match as given by $P$}
            \State continue
            \EndIf
            \State add $w$ to $S$ as a potential match for $u$
        \EndFor
    \EndFor
    \State \textbf{Return:} $S$
    \end{algorithmic}
\end{algorithm}

\subsection{Cost Function}\label{sec:costFunctions}

Once the list of start nodes in $G_T$ and their potential matches in $G_Q$ has been found, the algorithm performs a depth-first search in $G_T$ beginning at the start node with the fewest candidates in $G_Q$. During exploration, pairings are made between nodes using a modifiable cost function based on graph edit distance. Graph edit distance is a similarity measure between two graphs. Algorithms which compute graph edit distance find the edit path transforming one graph into the other with the fewest number of operations. The basic edit operations commonly used include node addition, node deletion, node substitution, edge addition, edge deletion, and edge substitution. We modify the graph edit distance functions used by Tu et. al. in \cite{tu2020inexact} to evaluate the cost of associating a pair of nodes during the depth-first search. We add a look-ahead component to assess how pairing $u \in V_T$ with $w 
\in V_Q$ may affect the matching in the future. \\

Let $f : V_T' \to V_Q'$ be a mapping from nodes $V_T' \subseteq V_T$ to nodes $V_Q' \subseteq V_Q$ (this may be a partial mapping). When considering a potential node pairing of $u \in V_T\setminus V_T'$ with a node $w \in V_Q \setminus V_Q'$, the function local\_cost is used, defined by

\begin{equation} 
\label{eq:node_cost}
\begin{split}
    \text{node\_cost}(u,w) = \gamma \cdot D_V(u, w) + (1-\gamma)\cdot L_V(u,w),
\end{split}
\end{equation}

\begin{equation} 
\label{eq:edge_cost}
\begin{split}
    \text{edge\_cost}(u,w) = \frac{\sum_{u' \in V_T'\setminus \{u\}} D_E((u,u'), (w, f(u')))}{N},
\end{split}
\end{equation}

\begin{equation} 
\label{eq:local_cost}
\begin{split}
        \text{local\_cost}(u,w) = \lambda_1 \cdot \text{node\_cost}(u,w) + \\
    (1-\lambda_1) \cdot \text{edge\_cost}(u,w).
\end{split}
\end{equation}

The user-set parameter $\lambda_1$ controls how much of the total pairing cost comes from node comparisons, and how much comes from edge comparisons. In Equation \ref{eq:node_cost}, $D_V$ is a function which calculates the average jaro-winkler score of node attributes that $u$ and $w$ share (or the node attributes to compare can be set by the user), and $L_V$ is our look-ahead function to assess future node pairings assuming $u$ gets paired with $w$ (defined below by Equation \ref{eq:look_ahead} below). The user-set parameter $\gamma$ controls the contributions of $D_V$ and $L_V$ to $\text{node\_cost}(u,w)$ (default value is $\gamma = 0.7$). In Equation \ref{eq:edge_cost}, $N$ is a normalizing constant, and the edge comparison function $D_E$, as defined by Tu et. al in \cite{tu2020inexact}, is given by 

\begin{eqnarray}
\label{eq:D_E}
D_E(e, f(e)) = \begin{cases}
            D(e, f(e)) \;\; \text{if $e \in E_T, f(e) \in E_Q$}\\
            D^-(e) \;\; \text{if $e \in E_T, f(e) \notin E_Q$}\\
            D^+(f(e)) \;\; \text{if $e \notin E_T, f(e) \in E_Q$}\\
            0 \;\; \text{if $e \notin E_T, f(e) \notin E_Q$}\\
        \end{cases}
\end{eqnarray}

where $e = (v_1,v_2)$ and $f(e) = (f(v_1), f(v_2))$. The functions $D$, $D^-$ and $D^+$ calculate edge attribute comparison, edge deletion, and edge addition costs, respectively. In our experiments, we set these functions equal to $1$ or $0$, but this is configurable. We use this to compare consistency within the mapped subgraphs by comparing edges in the induced subgraph on the nodes $V_T'$ in $G_T$, and the induced subgraph on the nodes $V_Q'$ in $G_Q$. With the edge functions $D, D^-$ and $D^+$ set to return values between $0$ and $1$, the total pairing cost for $u$ and $w$ has a minimum possible value of zero, which happens when $(u,w)$ is an exact match, and a maximum possible value of $1$ if $u$ and $w$ share no similarities in both node attributes and surrounding graph structure.\\

The graph edit distance functions are used to check consistency of the induced subgraph on the mapped nodes. In order to reduce the search space, we also check consistency within the unmapped nodes  via a function $L_V : V_T \times V_Q \to [0,1]$. This is a look ahead step, where we check to see if pairing node $u$ with node $w$ might produce a less than ideal matching in the future of the depth-first search. For a node $v$, let $v^1$ denote the number of unmapped neighbors of $v$, and let $v^2$ denote the number of unmapped 2-hop neighbors of $v$. We define the look-ahead function $L_V$ as

\begin{equation} 
\label{eq:look_ahead}
\begin{split}
        L_V(u,w) = 0.5 \cdot h(w^1 - u^1)/\text{deg}(w) + \\
        0.5 \cdot h(w^2 - u^2)/\text{2-deg}(w),
\end{split}
\end{equation}

where $\text{2-deg}(w)$ is the total number of 2-hop neighbors of $w$, and $h$ is given by

\begin{eqnarray*}
h(x) = 
\begin{cases}
    x \;\; \text{if $x \geq 0$}\\
    0 \;\; \text{else.}
\end{cases}
\end{eqnarray*}

Once a mapping $f : V_T' \to V_Q'$ is complete, the global cost of the mapping is calculated as the sum of the local costs, plus a penalty term for missing query nodes; these are given by

\begin{eqnarray}
\label{eq:cost}
    \text{cost}(f) = \left(\sum_{u \in V_T'} \text{local\_cost}(u, f(u))\right)\Big/|V_T'|,
\end{eqnarray}

\begin{eqnarray}
\label{eq:penalty}
    \text{missing\_node\_penalty}(f) = (|V_Q| - |V_Q'|)/|V_Q|,
\end{eqnarray}

\begin{equation} 
\label{eq:global_cost}
\begin{split}
        \text{global\_cost}(f) = \lambda_2 &\cdot \text{cost}(f) + \\
        (1 - \lambda_2) &\cdot \text{missing\_node\_penalty}(f),
\end{split}
\end{equation}

where $\lambda_2$ is a user-defined parameter (default value is $\lambda_2 = 0.7$). The penalty term for unmapped nodes in $G_Q$, given by Equation \ref{eq:penalty}, is added to the global cost of the mapping in order to map as many query nodes as possible. 

\subsection{Finding Mappings}
In order to find the optimal mapping, or multiple mappings, the algorithm ``backtracks" in the depth-first search for each start node in the target graph. We describe here what we mean by backtracking. During the initial depth-first search, the algorithm records node pairings that were acceptable matches but did not get paired. ``Acceptable matches" are determined by a user-set parameter which controls the maximum acceptable node pairing cost (note that this is set to zero to perform exact matching). The global cost of the mapping returned from this depth-first search is calculated as the sum of each individual node association cost, plus the penalty term for unmapped nodes. Backtracking then initiates, beginning part-way through the previous depth-first search at a node in the target graph that had another viable mapping option in the query graph. It then continues the depth-first search from this point, and returns the mapping it finds. If any pairing of nodes goes above the minimum global cost calculated for a mapping with the given start node so far, the depth-first search is terminated. This allows the algorithm to find the optimal mapping, since the best node pairing at each visited node may not return the optimal mapping overall. \\

Several techniques during the depth-first search are employed to reduce the search space. First, when looping through potential matches $w \in V_Q$ for node $u \in V_T$, it only considers nodes that are connected to nodes in the query graph that have been matched already. This reduces the number of cost calculations per visited node. Second, it calculates the maximum path length between the node in the query graph matched with the starting node, and any other node in the query graph. When adding neighbors of node $u \in V_T$ to the stack, it does not add neighbors that are farther than this distance from the starting node. This greatly reduces the nodes in the target graph visited in the depth-first search. Once all mappings from a given start node have been found, the algorithm may repeat the process with the next start node. Before starting this, it checks if the next start node has already been visited. If so, then this area of the graph has already been explored, and so it continues to the next start node. \\

\section{Experiments} \label{sec:experiments}
\subsection{Family Tree Graphs} \label{sec:familyTreeGraphs}

We used a dataset consisting of three family trees with the same surname, Adriaensen. These family trees are called Stamboom Adriaensen, Geneaology Adriaensen, and Family Tree Adriaensen. The graphs representing them have individuals as nodes, and an edge exists between two nodes if the individuals are parent/child or spouse/partner (relational information is included in the graph as edge attributes). Birth dates, death dates, and names were included in the graphs as node attributes. Three different subgraphs from the graph of Family Tree Adriaensen were manually selected as query graphs by viewing the graph in Cytoscape \cite{shannon2003cytoscape} and identifying nodes within a small area, representing a small family unit (we make sure that this gives a connected subgraph).   \\

The Adriaensen family tree graph data was collected from a free website called \href{https://www.genealogieonline.nl/en/}{Geneaology Online} where people can upload their own family tree information. The data was scraped from the website using Python and Beautiful Soup. Any private or hidden people were removed, and only individuals with public data were included in the dataset. Networkx graphs for each family tree were created, and the largest connected component was taken as the final graph. For all experiments, possible node pairings were given by a Jaro-Winkler score on name of at least $0.5$, while start nodes were taken to be nodes with exact match on name. During the depth-first search, all node attributes were compared in the node cost component of the local cost calculation. \\

\begin{table}[!ht]
\centering
\caption{{\bf Statistics for the family trees.}}
\begin{tabular}{||l c c c||}
\hline
& Stamboom & Geneaology & Family Tree \\
& (S) & (A) & (F)\\
\hline
\hline 
Number of Nodes & 1266 & 2543 & 2893\\
\hline
Number of Edges & 2087 & 4303 & 4912\\
\hline
\end{tabular}  
\label{tab:ft_target_graph_sizes}
\end{table}

\begin{table}[!ht]
\centering
\caption{{\bf Statistics for the chosen subgraphs of Graph $F$.}}
\begin{tabular}{||l c c c||}
\hline
& Query graph 1 & Query graph 2 & Query graph 3 \\
\hline
\hline 
Number of Nodes & 11 & 29 & 50\\
\hline
Number of Edges & 16 & 45 & 84\\
\hline
\end{tabular}  
\label{tab:ft_query_graph_sizes}
\end{table}

\begin{figure}[h]
\centering
\includegraphics[width=0.5\textwidth]{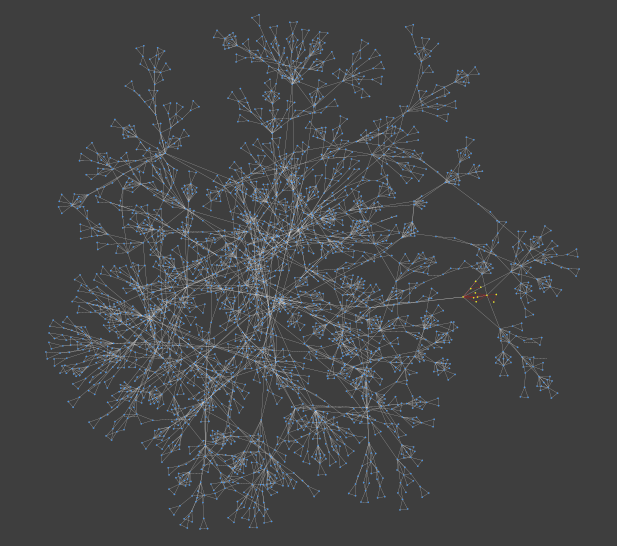}
\caption{{\bf Subgraph $Q_1$ in graph $F$.} Nodes of $Q_1$ highlighted in yellow, edges in red.}
\label{Q1_in_F}
\end{figure} 

\begin{figure}[h]
\centering
\caption{{\bf Exact matching: $Q_1$ in graph $S$.} Paired nodes in green, unmapped nodes of $Q_1$ in pink, unmapped nodes of $S$ in blue, unmapped edges in red.}
\includegraphics[width=0.5\textwidth]{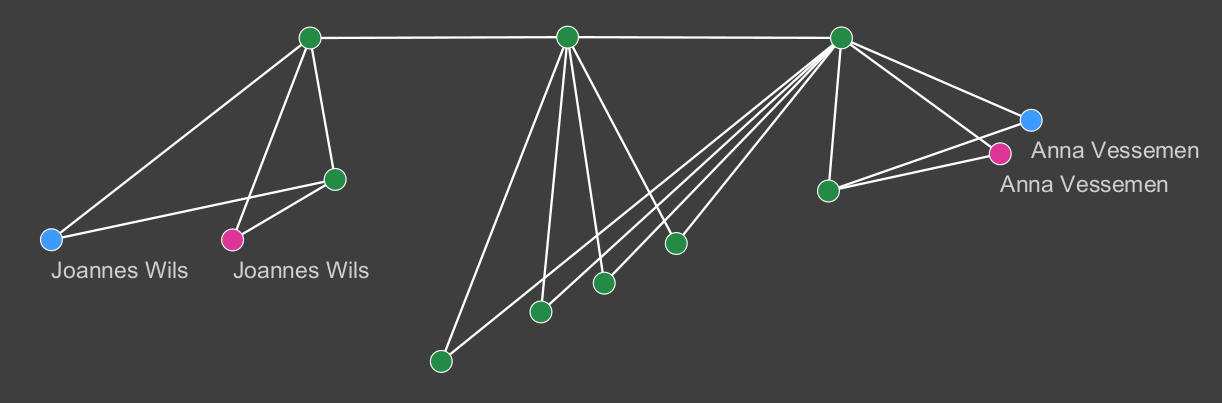}
\label{Q1_in_S_exact}
\end{figure}

We first consider query graph 1, which we call $Q_1$ (see Figure \ref{Q1_in_F}). All the nodes of $Q_1$ are present in all three graphs $A, F$ and $S$, so we expect the algorithm to return exact matchings (we check this before running the algorithm by the unique node labels, but we do not use the node labels in our experiments). The results are presented in Table \ref{tab:Q1_results}. When detecting $Q_1$ in graph $S$, not all of the nodes of the same name were paired during exact matching  (see Figure \ref{Q1_in_S_exact}). Edges between the unmapped nodes had different edges attributes (`spouse' versus `partner'). After enabling inexact matching these nodes do get paired.\\

\begin{figure*}[th]
\centering
\caption{{\bf Subgraph $Q_2$ in graph $F$.} Nodes of $Q_2$ highlighted in yellow, edges in red.}
\includegraphics[width=0.9\textwidth]{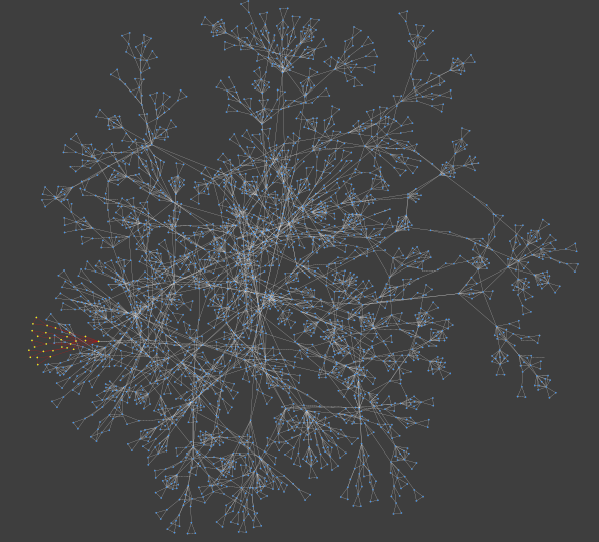}
\label{Q2_in_F} 
\end{figure*}

\begin{table}[!ht]
\centering
\caption{{\bf Detecting $Q_1$ in graphs $F, A$, and $S$.}}
\begin{tabular}{||ccccc||}
\hline
Target  & Runtime & Matching & Mapped nodes & Global\\
graph & (seconds) & type & in query graph & cost\\
\hline
\hline
$F$ & 0.011 & Exact & 11/11 & 0 \\
\hline
$A$ & 0.007 & Exact & 11/11 & 0 \\
\hline
$S$ & 0.09 & Exact & 9/11 & 0.068  \\
\hline
$S$ & 0.08 & Inexact & 11/11 & 0.042 \\
\hline
\end{tabular}
\label{tab:Q1_results}
\end{table}

\begin{figure}[h]
\centering
\caption{{\bf Inexact matching: $Q_2$ in graph $A$.} Paired nodes in green, unmapped node of $Q_2$ in pink, unmapped edges of $Q_2$ in red.}
\includegraphics[width=0.5\textwidth]{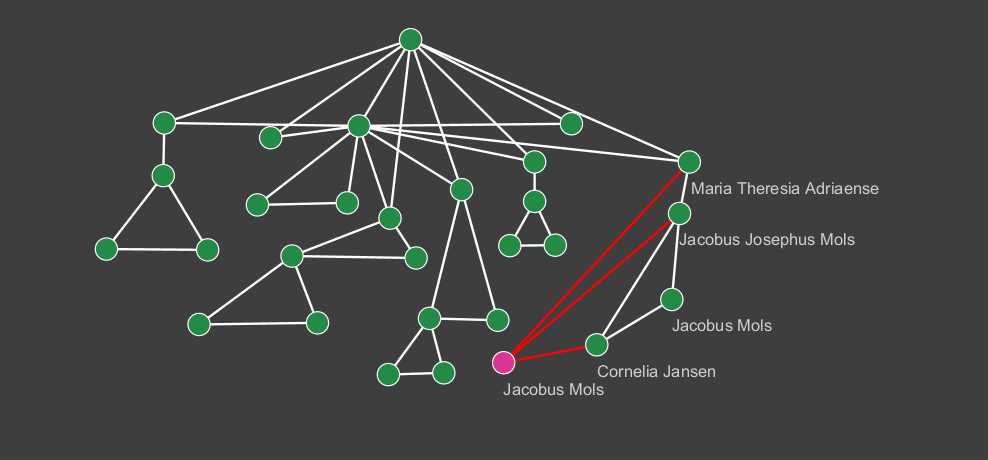}
\label{Q2_in_A_inexact} 
\end{figure}

\newpage
In $Q_2$, all the nodes are in graphs $F$ and $A$, none of the nodes are present in graph $S$ (see Figure \ref{Q2_in_F}).
Thus we expect exact matches in graphs $F$ and $A$.
Unmapped nodes, however, exist in the mapping returned from graph $A$ when only exact matching is enabled (see Table \ref{tab:Q2_results}).
The nodes with name Maria Theresia Adriaense did not get matched because they had different degrees (see Figure \ref{Q2_in_A_exact}). Because these did not get mapped, the algorithm did not explore the remaining unmapped nodes.
Once inexact matching is enabled, the algorithm correctly pairs the nodes with name Maria Theresia Adriense and explores the rest of the graph (see Figure \ref{Q2_in_A_inexact}).
The returned mapped has one unmapped node named Jacobus Mols, which is likely a duplicate entry. \\

\begin{table}[h]
\centering
\caption{{\bf Detecting $Q_2$ in graphs $F$ and $A$.}}
\begin{tabular}{||ccccc||}
\hline
Target  & Runtime & Matching & Mapped nodes & Global\\
graph & (seconds) & type & in query graph & cost\\
\hline
\hline
$F$ & 0.138 & Exact & 29/29 & 0 \\
\hline
$A$ & 0.01 & Exact & 24/29 & 0.05 \\
\hline
$A$ & 0.11 & Inexact & 28/29 & 0.01 \\
\hline
\end{tabular}
\label{tab:Q2_results}
\end{table}

\begin{figure}[h]
\centering
\caption{{\bf Exact matching: $Q_2$ in graph $A$.} Paired nodes in green, unmapped nodes of $Q_2$ in pink, unmapped nodes of $A$ in blue.}
\includegraphics[width=0.5\textwidth]{final_images/Q2_in_A_exact.png}
\label{Q2_in_A_exact}
\end{figure}

\begin{figure}[h]
\centering
\caption{{\bf Subgraph $Q_3$ in graph $F$.} Nodes of $Q_3$ highlighted in yellow, edges in red.}
\includegraphics[width=0.5\textwidth]{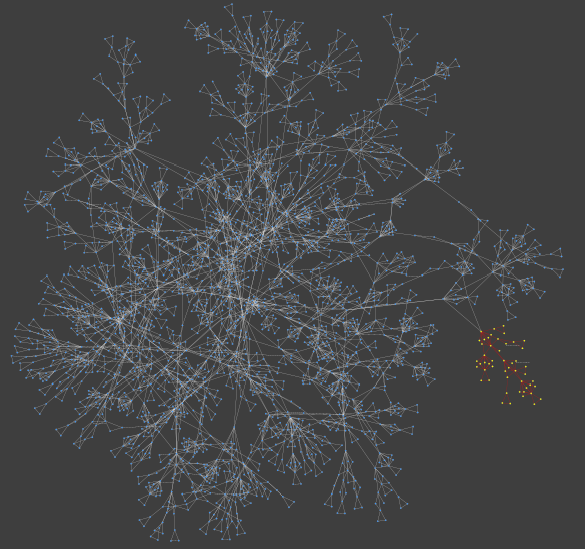}
\label{Q3_in_F}
\end{figure}

\begin{figure}[!h]
\centering
\caption{{\bf Exact matching: $Q_3$ in graph $A$.} Paired nodes in green, unmapped node of $Q_3$ in pink, unmapped node of $A$ in blue.}
\includegraphics[width=0.5\textwidth]{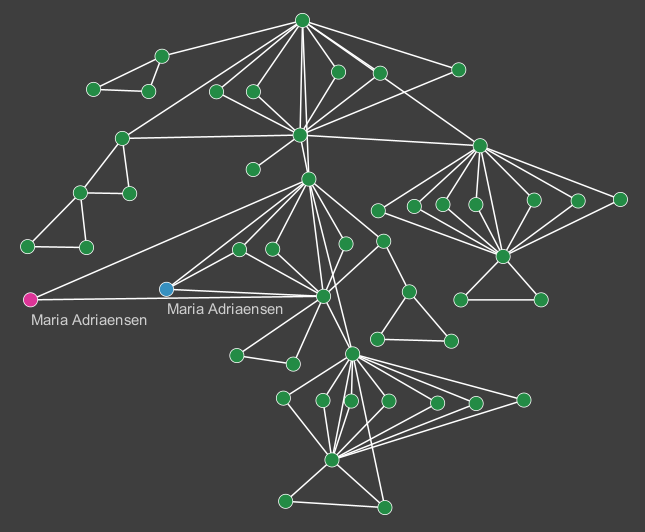}
\label{Q3_in_A_exact}
\end{figure}

\begin{figure}[!h]
\centering
\caption{{\bf Inexact matching: $Q_3$ in graph $A$.} Paired nodes in green, extra edge in graph $A$ in red.}
\includegraphics[width=0.5\textwidth]{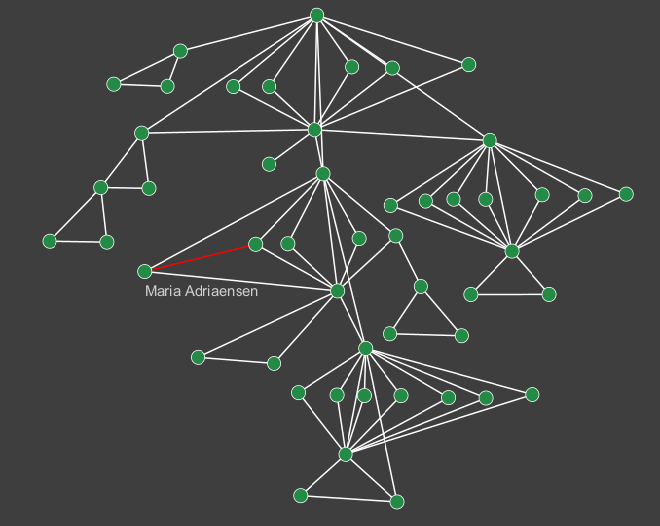}
\label{Q3_in_A_inexact}
\end{figure}

\begin{figure}[!h]
\centering
\caption{{\bf Exact matching: $Q_3$ in graph $S$.} Paired nodes in green, unmapped nodes of $Q_3$ in pink, unmapped nodes of $S$ in blue.}
\includegraphics[width=0.5\textwidth]{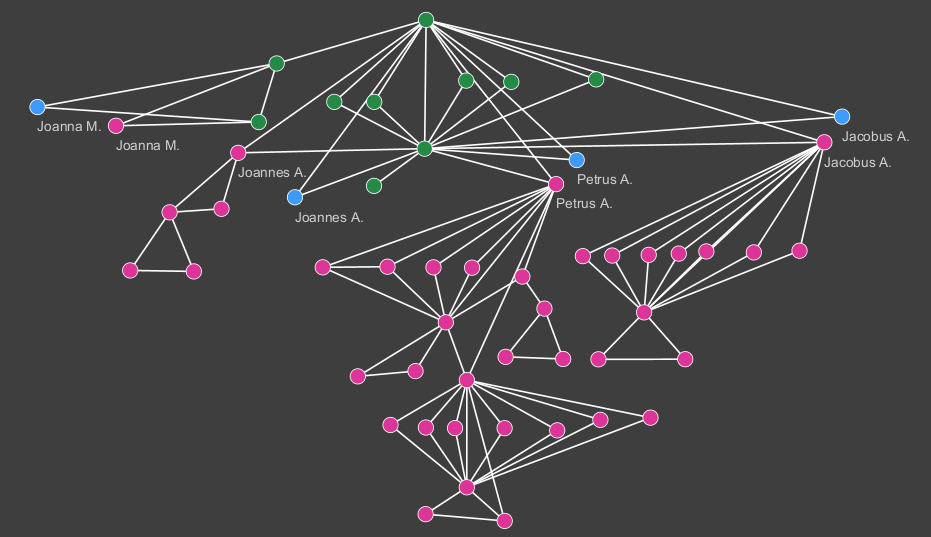}
\label{Q3_in_S_exact}
\end{figure}

\begin{figure}[!h]
\centering
\caption{{\bf Inexact matching: $Q_3$ in graph $S$.}  Paired nodes in green, unmapped nodes of $Q_3$ in pink.}
\includegraphics[width=0.5\textwidth]{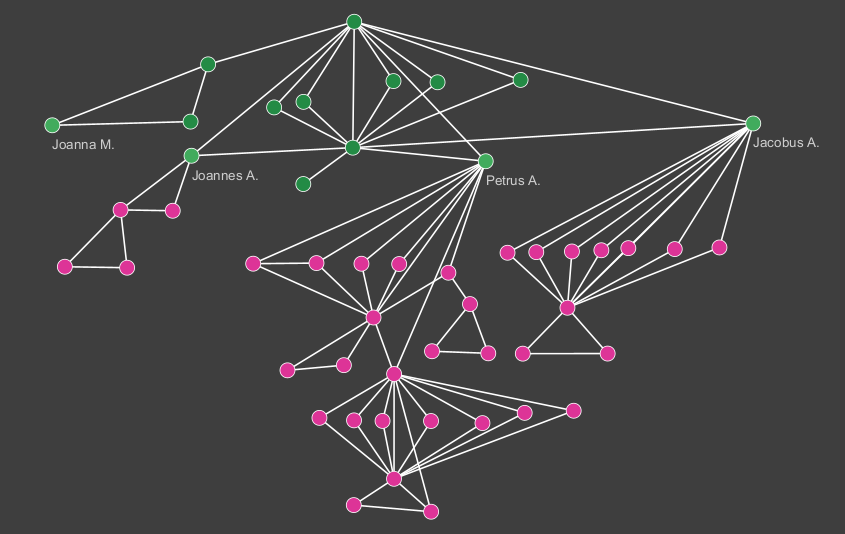}
\label{Q3_in_S_inexact}
\end{figure}

\newpage
In our last subgraph, $Q_3$, there are 14 nodes that are in all three graphs $A, F$, and $S$, and there are 36 nodes that are in graphs $F$ and $A$, not in graph $S$ (see Figure \ref{Q3_in_F}). Thus we expect the algorithm to pair all 50 nodes of $Q_3$ when detecting in graphs $F$ and $A$, and only 14 nodes when detecting in graph $S$. Results from our algorithm are given in Table \ref{tab:Q3_results}. The returned mapping from graph $A$ to $Q_3$ with only exact matching enabled has one unmapped node in $Q_3$ (see Figure \ref{Q3_in_A_exact}), which did not get mapped because of a difference in degree. Once inexact matching is enabled, the algorithm pairs all the nodes, leaving one edge between mapped nodes present in $G_T$, that is not present in $G_Q$ (see Figure \ref{Q3_in_A_inexact}). In Figure \ref{Q3_in_S_exact}, displaying the results of finding $Q_3$ in graph $S$, there are three unmapped nodes which match on node attributes, but their degree in $Q_3$ is higher than in $S$ and so they did not get mapped (nodes Jacobus A., Joannes A., and Petrus A.). The final unmapped node, Joanna M. matches on node attributes and degree, but is incident to an edge with attribute `spouse' in $S$ and `partner' in $Q_3$. With inexact matching enabled, the algorithm does map all 14 nodes present in both $S$ and $Q_3$ (see Figure \ref{Q3_in_S_inexact}), and does not map to the remaining nodes in $Q_3$ (the nodes in pink) that are not in graph $S$.\\

\begin{table}[h]
\centering
\caption{{\bf Detecting $Q_3$ in graphs $F, A$, and $S$.}}
\begin{tabular}{||ccccc||}
\hline
Target  & Runtime & Matching & Mapped nodes & Global\\
graph & (seconds) & type & in query graph & cost\\
\hline
\hline
$F$ & 0.31 & Exact & 50/50 & 0  \\
\hline
$A$ & 0.035 & Exact & 49/50 & 0.006 \\
\hline
$A$ & 0.177 & Inexact & 50/50 & 0.002 \\
\hline
$S$ & 0.015 & Exact & 10/50 & 0.24 \\
\hline
$S$ & 0.066 & Inexact & 14/50 & 0.23 \\
\hline
\end{tabular}
\label{tab:Q3_results}
\end{table}

The three subgraphs tested above were chosen to highlight potential common cases of noisy data that our algorithm can handle, such as differences in attributes and missing/extra edges or nodes. We first gave the results of exact matching for each subgraph to emphasize the errors in the matching graphs that our algorithm can manage. With $Q_1$, we see that our algorithm is capable of pairing the correct nodes even though the edge attributes differ. The results from $Q_2$ show that our algorithm correctly does not map extra nodes and edges in the query graph that do not exist in the target graph. Finally, with $Q_3$, we show that our algorithm can handle differences in degree between nodes that should be mapped (based on node attributes and surrounding graph structure). \\

\subsection{Function Identification in Binaries}

Another area where we can apply our algorithm is binary analysis, specifically for identifying function reuse in binaries. For C/C++ source code, many changes occur during compilation, and different compilers and optimization levels can produce very different binaries. This makes binary similarity calculations difficult and is one of the main challenges that algorithms performing this task must address \cite{david2017similarity}, \cite{feng2016scalable}, \cite{haq2021survey}. One technique used to execute this task given these challenges is inexact subgraph matching on a binary's control-flow graph \cite{dullien2005graph}. A control-flow graph (CFG) is a graph representing the possible paths that a binary file may traverse during its execution. Nodes in the control-flow graphs are basic blocks (a piece of code without any jumps or jump targets) and directed edges in the graph represent jumps in the control flow. We can identify specific functions then as subgraphs of a binary file’s control-flow graph. \\

We use data consisting of control-flow graphs of common use binaries generated by ANGR. To test exact matching and runtime, we choose the control-flow graph of a binary file as the target graph, and then choose functions within the binary with at least ten nodes as query graphs (see results in Table \ref{tab:exact_runtime_small}). We also record runtimes for query graphs of varying size in larger target graphs in Table \ref{tab:exact_runtime_large}. From these tables, we can see that in target graphs up to 38,000 nodes, we have most runtimes for query graphs ranging from 18 to 114 nodes around one second, with a few exceptions. In some cases, depending on the graph structure, the algorithm may be slower. For example, if a graph has a lot of leaf nodes with similar node attributes this can increase runtime.


\begin{table}[!ht]
\centering
\caption{{\bf Exact matching results.}}
\begin{tabular}{||c c c c||}
\hline
Function & Query graph size& Runtime & \# mapped nodes\\
name & (\# nodes, \# edges )& (secs) & in query graph\\
\hline
\hline
die & (16, 16) & 0.08 & 16/16 \\
\hline
usage & (19, 10) & 0.004 & 19/19\\
\hline
mbiter\_multi\_next & (26, 32) & 0.005 & 26/26\\
\hline
version\_etc\_arm & (42, 50) & 0.181 & 38/42\\
\hline
trim2 & (68, 100) & 0.041 & 68/68\\
\hline
mbsstr\_trimmed & (79, 117) & 0.178 & 79/79 \\
\hline
knuth\_morris & (85, 127) & 0.03 & 85/85 \\
\hline
mbsstr & (118, 183) & 0.07 & 118/118\\
\hline
main & (133, 197) & 0.66 & 131/133\\
\hline
\end{tabular}  
\label{tab:exact_runtime_small}
\begin{flushleft}
Target graph cmp-gcc-x86-64-O0, size = (1765, 3111)
\end{flushleft}
\end{table}

\begin{table}[!ht]
\centering
\caption{\bf Runtime in seconds for exact matching on larger target graphs.}
\resizebox{.45\textwidth}{!}{
\begin{tabular}{|ccccccccc|}
\hline
 & \multicolumn{8}{c|}{Number of query nodes} \\ 
 \hline
 Target&&&&&&&&\\
 graph& 18 & 24 & 32 & 45 & 54 & 68 & 83 & 114 \\
 size &&&&&&&&\\
\hline
\hline
(9555,  & 0.023 & 0.018 & 0.025 & 0.021 & 0.042 & 0.113 & 0.281 & 0.333\\
 18273)&&&&&&&&\\
\hline
(15107, & 0.127 & 0.195 & 0.129 & 0.159 & 0.991 & 0.313 & 4.634 & 1.488 \\
 34706)&&&&&&&&\\
\hline
(15115, & 0.917 & 0.043 & 0.262 & 0.121 & 0.106 & 0.151 & 0.145 & 0.316 \\
 28097)&&&&&&&&\\
\hline
(26084, & 0.141 & 1.51 & 0.24 & 1.766 & 3.065 & 2.087 & 2.285 & 3.135 \\
 61180)&&&&&&&&\\
\hline
(38117, & 2.284 & 0.064 & 1.336 & 1.53 & 0.12 & 1.305 & 2.19 & 2.042 \\
 65936)&&&&&&&&\\
\hline
(66098, & 0.257 & 0.191 & 8.041 & 0.293 & 0.239 & 4.853 & 5.493 & 9.206 \\
 134423)&&&&&&&&\\
\hline
\end{tabular}  
}
\label{tab:exact_runtime_large}
\end{table}

Note that in Table \ref{tab:exact_runtime_small}, we see that the algorithm sometimes returns incomplete mappings, even though there exists an exact match.  In these cases, there is a pairing $(u,w)$ made such that $u \neq w$, but $\text{local\_cost}(u,w) = 0$. Thus $u$ is mapped incorrectly. Then later in the depth-first search when the node in $G_T$ that should be mapped to $w \in G_Q$ is reached, $w$ is no longer an option to be mapped to, and so this creates another incorrect pairing, and so on. \\

\begin{figure}[!h]
\centering
\caption{{\bf block\_read.} Mapped nodes are color-coded, red nodes are unmapped. A: block\_read in -O1 (target graph). B: block\_read in -O2 (query graph).}
\includegraphics[width=0.48\textwidth]{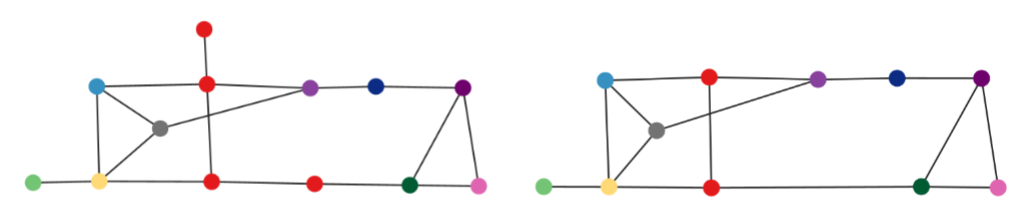}
\label{block_read}
\end{figure}

\begin{figure}[!h]
\centering
\caption{{\bf set\_program\_name.} Mapped nodes are color-coded, red nodes are unmapped. A: set\_program\_name in -O1 (target graph). B: set\_program\_name in -O2 (query graph).}
\includegraphics[width=0.48\textwidth]{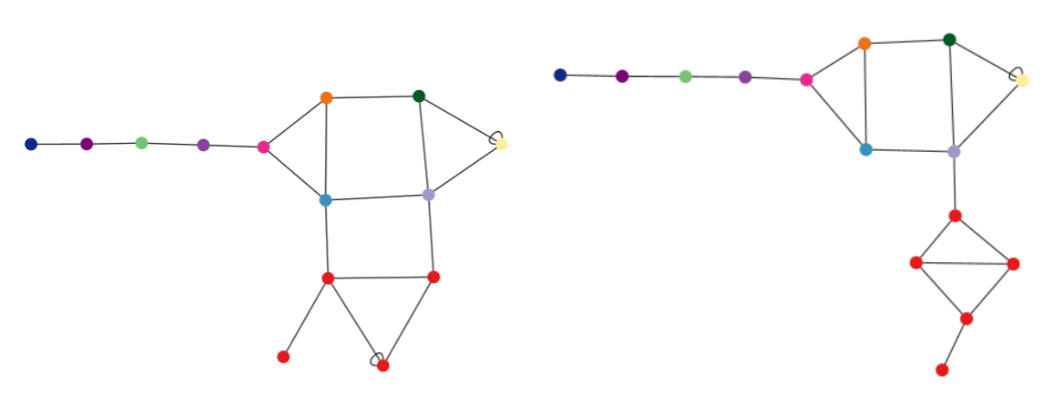}
\label{set_program_name}
\end{figure}

\begin{figure}[!h]
\centering
\caption{{\bf freea.} Mapped nodes are color-coded, red nodes are unmapped. A: freea in -O1 (target graph). B: freea in -O2 (query graph).}
\includegraphics[width=0.48\textwidth]{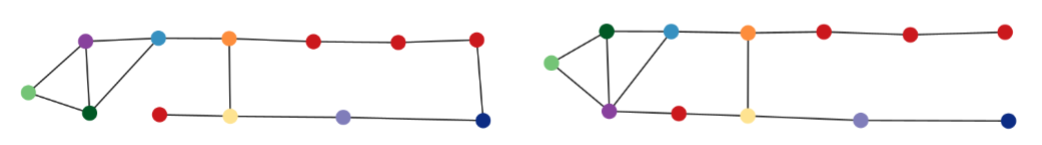}
\label{freea}
\end{figure}

To test inexact matching, we choose a control-flow graph for a binary file compiled with optimization level 1 for the target graph. Query graphs are chosen to be functions in the same binary file, compiled with the same compiler, but compiled with optimization level 2. The target graph is labeled cmp-gcc-x86-O1, and the query graphs come from the graph labeled cmp-gcc-x86-O2. Three query graphs are tested, given by the function names set\_program\_name, block\_read, and freea. We expect the functions in cmp-gcc-x86-O1 and cmp-gcc-x86-O2 to be similar, but not isomorphic (see results in Table \ref{tab:inexact_results}). In Figures \ref{block_read}, \ref{set_program_name}, and \ref{freea}, representing the mappings returned from our algorithm for each query graph, we note that there are several unmapped nodes (the red nodes) in each query graph. Based on graph structure alone, these nodes appear as though they should have been mapped. These unmapped nodes however are dissimilar in node attributes, which is why they were not mapped. From these results, we conclude that our algorithm successfully identified the query graphs, given that the node attributes for some of the nodes between the two graphs were quite varied. \\

\begin{table}[!ht]
\centering
\caption{{\bf Inexact matching results.}}
\begin{tabular}{||c c c c c||}
\hline
 & Query graph &  & \# mapped  & \\
Function & size  & Runtime & nodes  & Global\\
name & (\# nodes, & (secs) & in & Cost\\
 &  \# edges )&  &  graph& \\

\hline
\hline
set\_program\_name & (15, 20) & 0.151 & 10/15 & 0.13 \\
\hline
block\_read & (11, 15) & 0.209  & 9/11 & 0.12 \\
\hline
freea & (12, 14) & 0.087 & 8/12  & 0.15 \\
\hline
\end{tabular}  
\begin{flushleft}
  Target graph cmp-gcc-x86-64-O1, size = (1493, 2719).  
\end{flushleft}
\label{tab:inexact_results}
\end{table}

\begin{figure}[!h]
\centering
\caption{{\bf Fuzzy matching.} F1-score for fuzzy subgraphs.}
\includegraphics[width = 0.5\textwidth]{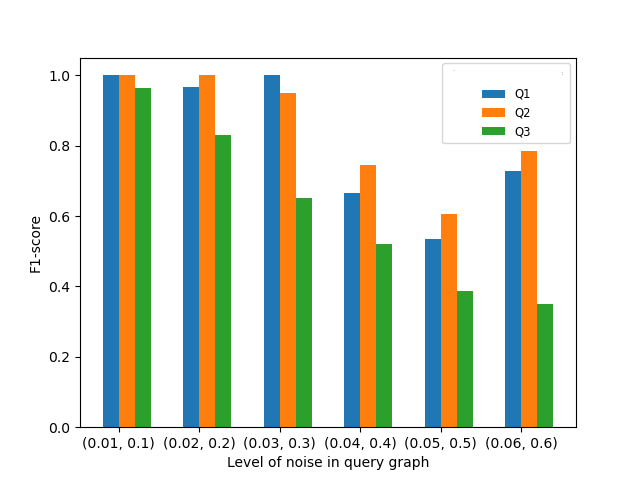}
\label{fuzzy_results}
\end{figure}

Finally, we test the effectiveness of our algorithm. We pull subgraphs from the control-flow graph cmp-gcc-x86-O0 to use as query graphs, and add noise in varying levels to both the graph structure and the node attributes.
With this approach, we have ground truth labels on the node pairings that we use to calculate the f1-score of the mapping returned by the algorithm (note that we do not use the node labels for pairing).
To add noise to the query graphs, first we modify the graph structure (see Table \ref{tab:noise_level_graph_structure}), and then we modify the node attributes (see Table \ref{tab:noise_level_node_attributes}).
For the modifications to graph structure, the noise level gives the probability of deleting each node (and its adjacent edges), deleting each edge, and adding an edge between any two nodes (the probability for edge addition is divided by four, so that we have a proportional number of added edges). For the modifications to node attributes, the noise level gives the probability of altering node attributes for each node. The number of attributes to change, and which attributes to change are chosen uniformly at random. Results from our algorithm given these fuzzy subgraphs are given in Figure \ref{fuzzy_results}.
As expected, the performance of the model decreases as the noise level increases. We note that our algorithm performs better on $S_1$ and $S_2$, the smaller subgraphs, than on $S_3$. For $S_3$, we have an f1-score of above 0.8 up to a noise level of 0.02 on graph structure and 0.2 on node attributes.
For $S_1$ and $S_2$, we have an f1-score of above 0.8 up to a noise level of 0.03 on graph structure, and 0.3 on node attributes.
We conclude that our algorithm is most effective for smaller query graphs, as it can successfully locate the correct node pairings even with significant changes in graph structure and node attributes. \\

\begin{table}[!ht]
\centering
\caption{{\bf Noise level in query graph when modifying graph structure.}}
\begin{tabular}{||c | c | c c c c c c||}
\hline
\multicolumn{2}{|c}{Noise Level} & 0.01 & 0.02 & 0.03 & 0.04 & 0.05 & 0.06 \\
\hline
\hline

$S_1$& \# nodes removed& 0 & 0 & 0 & 0 & 1 &1 \\
(16,16) &\# edges removed & 0 & 0 & 1 & 1& 2 & 3\\
&\# edges add & 0 & 0 & 1 & 3 & 2 & 2\\
\hline
$S_2$& \# nodes removed& 0 & 0 & 1& 1 & 2 &2 \\
(32,48) &\# edges removed & 1 & 1 & 5 & 2& 6 & 6\\
&\# edges add & 1 & 2 & 3 & 6 & 10 & 7\\
\hline
$S_3$& \# nodes removed& 0 & 2 & 2 & 4 & 3 &3 \\
(55,77) &\# edges removed & 0 & 9 & 5 & 10& 10 & 16\\
&\# edges add & 4 & 7 & 8 & 15 &16 & 18\\
\hline
\end{tabular}  
\label{tab:noise_level_graph_structure}
\end{table}

\begin{table}[!ht]
\centering
\caption{{\bf Noise level in query graph when modifying node attributes.}}
\begin{tabular}{||c | c | c c c c c c||}
\hline
\multicolumn{2}{|c}{Noise Level} & 0.1 & 0.2 & 0.3 & 0.4 & 0.5 & 0.6 \\
\hline
\hline
$S_1$ & \# nodes with  & 6 & 6 & 9 & 10 & 6 & 11 \\
(16, 16)  & attributes altered  &  &  &  & &  &  \\
\hline
$S_2$ & \# nodes with  & 8 & 9 & 12 & 20 & 19 & 21 \\
(32, 48) & attributes altered  &  &  &  & &  &  \\
\hline
$S_3$ & \# nodes with  & 12 & 19 & 30 & 33 & 30 & 28 \\
(55, 73) & attributes altered  &  &  &  & &  &  \\
\hline
\end{tabular}  
\label{tab:noise_level_node_attributes}
\end{table}

\section{Conclusion}\label{sec:conclusion}
We proposed an algorithm for inexact subgraph matching for graphs with node and edge attributes. The algorithm is capable of finding exact matches if they exist, and finds the best inexact match otherwise. Use of a highly modifiable cost function for pairing nodes allows the user to set the importance of node attribute versus graph structure in the mapping. Results of our algorithm on family tree graphs and control-flow graphs, 
with the tasks of identifying family units and function reuse in binaries respectively, show the effectiveness and scope of our algorithm. We conclude from these experiments that our algorithm is most effective on small query graphs, even ones with considerable noise, and is efficient on target graphs up to about 60,000 nodes. We recommend our algorithm in the case when the user is dealing with smaller graphs, and requires high flexibility in the types of query graphs to be located or in the types of datasets, or if the node attributes in the graphs are varied.  

\section{Future Work}\label{sec:future_work}
We propose several directions for future work, including reducing runtime, adding modifications to allow for more variation in the returned subgraph, and extending the application of identifying function re-use in binaries to third-party library detection. Our algorithm currently doesn't allow for node gaps, that is, an unmapped node in between two mapped nodes. This is easily altered however. During the depth-first search, candidate nodes are those given by a node attribute comparison which are also connected to a mapped node. Expanding this to nodes that are two hops away from a mapped node would allow for node gaps. If the query graph contains a lot of noise compared to the target graph, this may be a beneficial approach, however if not, it may increase runtime without offering much improvement to the mapping. \\


The application of this algorithm can be extended to third-party library (TPL) detection in binaries. The goal of TPL detection is as follows: determine if and where a candidate binary (the third-party library) is used in a target binary (or binaries). One approach for this task is incorporating multidimensional features and analyzing control-flow graphs or function control graphs (FCG) of the binaries. In the function control graph of a binary, the nodes are subroutines (such as methods, functions), and edges between nodes represent the caller-called relationship between two subroutines. The very general approach for methods which use CFG or FCG analysis is to use basic features of functions (instructions and library calls), and another type of comparison (multidimensional feature vectors for example) to make matches between nodes in the function control graphs representing the target binary and a candidate binary. First, anchor nodes are identified (nodes which match exactly, on the basic features for example), then areas of the FCGs are explored to make further matches (based on multidimensional feature vectors for example). Based on these matches, a final similarity score between the FCGs is calculated, which determines whether the candidate binary was used in the target binary or not. Our algorithm can already perform function re-use detection between binaries, and thus could potentially be used to make the initial matches between nodes in the FCGs, as well as for exploring the FCG. The global cost of the resulting matching subgraph in the FCG can be used to evaluate whether a third-party library was used in the binary or not.\\

\section*{Acknowledgment}
The authors wish to thank the Department of Energy (DOE) Cybersecurity, Energy Security, and Emergency Response (CESER) and the Cyber Testing and Resilience of Industrial Control Systems (CyTRICS) Program including Idaho National Laboratory (INL), Lawrence Livermore National Laboratory (LLNL), National Renewable Energy Laboratory (NREL), Oakridge National Laboratory (ORNL), and Sandia National Laboratory (SNL).
They also acknowledge Doug Dennis and Tyler Williams of PNNL for the control-flow graph dataset.

\end{document}